\documentclass{article}
\usepackage{epsfig}

\title{Photon number conservation and photon interference}

\author{M\' aty\' as Koniorczyk$^{a,b}$ and J\' ozsef Janszky$^{a,b}$
\\
$^{a}$Department of Nonlinear and Quantum Optics,\\ Research Institute for
  Solid State Physics and Optics, \\ Hungarian Academy of Sciences, \\ P.O.
  Box 49. H-1525 Budapest, Hungary \\
$^{b}$
  Institute of Physics, University of P\' ecs, \\
  Ifj\' us\' ag \' ut 6, H-7624 P\'ecs, Hungary}

\date{October 20, 2001.}

\begin{document}
\maketitle
\begin{abstract}
  The group theoretical aspect of the description of passive lossless optical
  four-ports (beam splitters) is revisited. It is shown through an example,
  that this approach can be useful in understanding interferometric schemes
  where a low number of photons interfere. The formalism is extended to
  passive lossless optical six-ports, their $SU(3)$-theory is outlined.
\end{abstract}

\section{INTRODUCTION}

Group theoretical methods are prevalent in the physics of the 20th
century. They have found their applications in the field of quantum optics as
well. One of the typical application is related with nonclassical states of
light~\cite{josab2_458,pra53_1886,optexpr8_76,pra60_R1737}. Another
application is involved in the description linear and nonlinear optical
multiports~\cite{qsemicl11_571,pra62_033808,joptb2_133}.

In this paper we will consider linear optical devices, such as beam splitters
and tritters (i.e. optical six-ports). These are basic devices of
interferometry, and building blocks of several schemes that have attracted
much attention recently, such as polarization
teleportation~\cite{nature390_575}, quantum lithography~\cite{prl85_27331}, or
quantum computation with linear optics~\cite{nature409_46}. In such
applications, states with a low number of photons interfere on these devices.

The operation of a lossless beam splitter from quantum mechanical
point of view was investigated by several authors. A thorough summary
of this topic was given by Campos, Saleh and Teich~\cite{pra40_1371},
who emphasize the group theoretical aspects of the description, which
are originating from the conservation of photon number. The
description relies on the Schwinger-representation of $su(2)$
Lie-algebra, i.e. angular momenta. In Section~\ref{sect:splitters} we
give a brief summary of these methods. In Section~\ref{sect:scissors}
we show an application, namely the generalized ``quantum scissors''
device~\cite{prl81_1604,pra60_4965,pra62_013802}, which is also
teleporter for photon-number state superpositions. We demonstrate how
the $su(2)$ algebra and the properties of the beam splitter
transformation can help us in understanding the operation of a device
of photon interference.

Tritters, i.e. three-input three-output passive linear optical
elements also have found several applications, such as in quantum
homodyning~\cite{pra54_856}, entanglement realization~\cite{pra55_2564},
Bell-experiments~\cite{pra59_3200}, and optical realization of certain
nonunitary transformations~\cite{jmodopt47_487}. The analogous
treatment to that of beam-splitters, i.e. application of $SU(3)$
symmetry in the description of such devices has not yet been
emphasized. In Section~\ref{sect:tritters} we outline the appropriate
formalism. In Section~\ref{sect:concl} we summarize the conclusions.

\section{BEAM SPLITTERS REVISITED}
\label{sect:splitters}

A beam splitter, or linear coupler is a device having two input and two output
ports, each of which are single modes of the electromagnetic field. Let $\hat
a_1$ and $\hat a_2$ denote the annihilation operators of the input ports, and
$\hat b_1$ and $\hat b_2$ those of the outputs respectively. In a passive
lossless linear beam splitter (or in a linear coupler) these are connected via
\begin{equation}
  \label{eq:bsdef}
  \hat b_i=\sum_{j=1}^{2} U_{ij} \hat a_j, \quad i=1,2 \quad U\in SU(2),
\end{equation}
thus the $2\times 2$ matrix $U$ is a matrix corresponding to the fundamental
representation of $SU(2)$. The details of theory of lossless passive beam
splitters can be found in Ref.~\cite{pra40_1371}. Our task is now to emphasize
the strictly group-theoretical aspects of the theory.

Both the input and the output pair of modes can be regarded as two-dimensional
oscillators, possessing $SU(2)$ symmetry. According to Schwinger
representation of angular momenta\cite{Biedenharn_Louck}, the generators 
$\hat L_1,\hat L_2, \hat L_3$ of the $su(2)$ Lie-algebra can be constructed as
\begin{equation}
  \label{eq:su2_gen}
  \hat L_k=\matrix{(\hat a_1^\dag &\hat a_2^\dag )\cr & }
\left( \matrix{  &\frac{1}{2}\hat \sigma _k& \cr & }\right)
\left(\matrix{\hat a_1\cr \hat a_2}\right), \quad k=1,2,3\, ,
\end{equation} 
where the $\hat \sigma_k$-s are the Pauli-matrices.  The output operators
$\hat b_i$ realize the $su(2)$ Lie-algebra in the same way, these generators
will be denoted by $\hat K_1,\hat K_2, \hat K_3$ The consequence of this is,
that the two-mode number states $|n,m\rangle$, $n,m\in {\bf n}$ can be divided
into $SU(2)$-multiplets.  We consider input states, the method is the same for
the output states.  

One may construct the operator  
\begin{equation}
  \label{eq:lop}
\hat l=\frac{1}{2} (\hat a_1^\dag \hat a_1+\hat a_2^\dag \hat a_2), 
\end{equation}
from which the operator of the ``square of angular momentum'', the
is the Casimir-operator of the algebra, can be constructed as
\begin{equation}
  \label{eq:lsquare}
  \hat L^2 =\hat l (\hat l + \hat 1).
\end{equation}
The multiplets can be indexed by the eigenvalue of the Casimir-operator. In
the theory of the angular momenta, it is usual to use the eigenvalue of $\hat
l$ instead, as it is in a one-to-one correspondence with the eigenvalue of
$\hat L^2$. In our case, multiplet of index $l$ is the set of the number
states with $2l=n+m$.  The states in a multiplet are indexed by the
eigenvalues $l_3=\frac{1}{2}(n-m)$ of $\hat L_3$.  Thus instead of the photon
number, states can be indexed as
\begin{equation}
  \label{eq:su2ind}
  |n,m\rangle=|l,l_3\rangle.
\end{equation}
The ladder operators $L_+=\hat a_1^\dag\hat a_2$, $L_-=\hat a_2^\dag\hat a_1$
defined in the standard way can be applied to increase and decrease the index
$l_3$. The same relabelling of states can be defined for the number states at
the output.

The beam splitter itself is also an $SU(2)$ device according to
Eq.~(\ref{eq:bsdef}). There are two important consequences of this fact. One
of these is, that the Lie-algebras at the input and at the output are related
as
\begin{equation}
  \label{eq:o3}
  \hat K_k=\sum_{l=1}^3 {\cal O}_{kl}\hat L_l, \quad k=1,2,3\, ,
\end{equation}
where ${\cal O}$ is the element of $SO(3)$, rotations of the
three-dimensional real vector-space, corresponding to $U$ in
Eq.~(\ref{eq:bsdef}). This provides us with the opportunity of
visualizing the action of the device as a rotation of a vector in the
three-dimensional space. The detailed analysis can be found in
Ref.~\cite{pra40_1371}.  The other important consequence is, that
the multiplets of the states are invariant subspaces of the beam
splitter transformation in Eq.~(\ref{eq:bsdef}), namely, $l$ is
conserved by the transformation. Thus the notation in
Eq.~(\ref{eq:su2ind}) is very suitable in the description of the beam
splitter transformation.

We have seen, that there are three direct consequences of the $SU(2)$
symmetry: relabeling of states, conservation of $l$, and the
connection with $SO(3)$. In the next section we present a simple
application of few-photon interference, where the multiplet way of
thinking proves to be useful.

\section{AN APPLICATION OF $SU(2)$ SYMMETRY}
\label{sect:scissors}

The quantum scissors device\cite{prl81_1604,pra60_4965,pra62_013802} is a tool
for quantum state design of running wave states, exploiting quantum
nonlocality. Under certain conditions, it can be regarded as a teleporter for
photon number states. We examine some aspects of the latter case in detail.
\begin{figure}[htbp]
  \begin{center}
   \epsfig{file=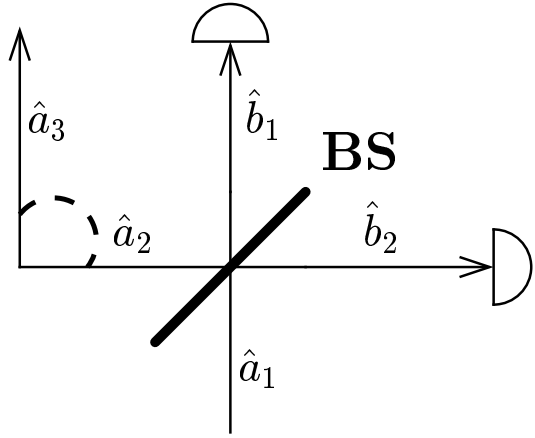}
\caption[scissors]
{ \label{fig:scissors} 
The ``quantum scissors'' device analyzed in Section\ref{sect:scissors}. The
annihilation operators of the modes in arguments are shown in the picture.
}  
  \end{center}
\end{figure}
Consider the scheme in Fig.~\ref{fig:scissors}. The aim is the conditional
teleportation of the state 
\begin{equation}
  \label{eq:scissin}
  |\Psi_{\mbox{\scriptsize{in}}}\rangle
=A_0|0\rangle_1+A_1|1\rangle_1+A_2|2\rangle_1
\end{equation}
in mode 1, which is at Alice, to Bob at mode 3. Alice and Bob share an
entangled state
\begin{equation}
  \label{eq:epr}
  |\Psi_{\mbox{\scriptsize{EPR}}}\rangle=C_{-1}|2\rangle_2 |0\rangle_3+
C_{0}|1\rangle_2 |1\rangle_3+
C_{1}|0\rangle_2 |2\rangle_3
\end{equation}
in modes $2$ and $3$ (we have applied $SU(2)$ notation in the indices
of $C$ coefficients). Modes $1$ and $2$ interfere on the beam splitter
BS. The output ports of the beam splitter are incident on detectors,
which are supposed to be ideal photon counters realizing projective
quantum measurement on the number state basis. The teleportation is
successful, if the detectors count one single photon each, in
coincidence.

The measurement can be described as follows. The detection of the
photons is the annihilation of two photons to vacuum, described by the
operator $\hat b_1\hat b_2$. Therefore, in order to obtain the
teleported state, we write the state of all three modes
$|\Psi_{\mbox{\scriptsize{m}}}\rangle_{123}$ after the action on the
beam splitter into the form
\begin{equation}
  \label{eq:psim}
 |\Psi_{\mbox{\scriptsize{m}}}\rangle_{123}= \hat A^\dag(\hat b_1^\dag
  ,\hat b_2^\dag ,\hat a_3^\dag )|0\rangle,
\end{equation}
where the operator $\hat A^\dag(\hat b_1^\dag ,\hat b_2^\dag ,\hat a_3^\dag)$
  is a polynomial of the creation operators, and generates
  $|\Psi_m\rangle_{123}$ from the vacuum. The projection by the measurement
  drops all the summands in the expression of $\hat A^\dag(\hat b_1^\dag
  ,\hat b_2^\dag ,\hat a_3^\dag)$, except for that containing $\hat b_1^\dag
  \hat b_2^\dag $, thus the resulting (teleported) state can be read out from
  the expression of $\hat A^\dag(\hat b_1^\dag ,\hat b_2^\dag ,\hat
  a_3^\dag)$.  As $\hat b_1$ and $\hat b_2$ originate from a beam splitter
  transformation, it is worth collecting them into $SU_2$ multiplets.  Let us
  introduce the notation
\begin{equation}
^{2l}\hat M_{l_3}=\hat b_2^{\dag l+l_3}\hat b_1^{\dag l-l_3}, \quad
l_3=-l\ldots l 
\label{eq:meas_op}
\end{equation} 
for the outcome operators. These are groups of creation operators indexed by
the $SU(2)$ indices.
Furthermore, given a set of arbitrary
coefficients ${^{2l}{\cal A}}_{l_3},\ l_3=-l\ldots l$, let there be
\begin{equation}
^{2l}\hat{\cal M}_{\cal A}=\sum_{l_3=-l}^l 
{^{2l}{\cal A}}_{l_3} {^{2l}\hat M_{l_3}}
\label{eq:multiplet}
\end{equation}
a linear combination of outcome operators in the $l$-th multiplet, with
coefficients ${^{2l}{\cal A}}$. Different calligraphic letters in the index
shall mean a different set of parameters in this notation.

Notice, that maximum of $4$ photons can be present at the beam splitter. The
coefficients $A_0,A_1,A_2$ from Eq.~(\ref{eq:scissin}) should appear in $\hat
A^\dag(\hat b_1^\dag ,\hat b_2^\dag ,\hat a_3^\dag)$ before $\hat a_3^\dag$ on
powers determined by Eq.~(\ref{eq:scissin}), due to the linearity of the
system. Thus in general we have 
\begin{eqnarray}
\hat A^\dag (\hat b_1^\dag ,\hat b_2^\dag ,\hat a_3^\dag)=&&
A_0({^{2}\hat{\cal M}}_{\cal A}+{^{1}\hat{\cal M}}_{\cal B} \hat
a_3^\dag +{^{0}\hat{\cal M}}_{\cal C} \hat a_3^{\dag2}) \nonumber \\
+&&A_1({^{3}\hat{\cal M}}_{\cal D}+{^{2}\hat{\cal M}}_{\cal E}
\hat a_3^\dag +{^{1}\hat{\cal M}}_{\cal F} \hat a_3^{\dag2}) \nonumber \\
+&&\frac{A_2}{\sqrt{2}}({^{4}\hat{\cal M}}_{\cal G} +{^{3}\hat{\cal
M}}_{\cal H} \hat a_3^\dag +{^{2}\hat{\cal M}}_{\cal I} \hat a
_3^{\dag2}).
\label{eq:str}
\end{eqnarray}
The coefficients $C$ of the entangled state in Eq.~(\ref{eq:epr}), and the
parameters of the unitary operator $U$ describing the beam splitter BS are
included in the coefficients denoted by calligraphic letters.

It can be noticed that the multiplet structure suggested by the nature of the
beam splitter transformation is reflected in the structure of the operator
creating the output state.  Only the outcomes in the $^{2}\hat{\cal M}_{\cal
A},^{2}\hat{\cal M}_{\cal E},^{2}\hat {\cal M}_{\cal I}$ multiplets appear with all
three $A$ coefficients of the input state. Only the outcomes in these
multiplets can provide teleportation, since the state obtained after the
measurement on mode $\hat a_3$ depends on all three A coefficients. In
the case of a measurement outcome corresponding to an other multiplet some of
the information is lost.  The whole information is transferred if the total
number of detected photons is 2.

The ${^{2l}{\cal A}}$,~${^{2l}{\cal E}}$, and~${^{2l}{\cal I}}$ coefficients
depend on the beam splitter parameters and the $C$ parameters of the entangled
state in Eq.~(\ref{eq:epr}). It is possible to set these parameters so that
${^2{\cal A}}_0={^2{\cal E}}_0={^2{\cal I}}_0=1/3$. The actual determination
of the appropriate entangled state and beam splitter is a geometrical problem
discussed in detail in Ref.~\cite{pra62_013802}. In the case of measurement
outcome described by ${^{2}\hat M_0}$, i.e. ~detection one photon on both
detectors in coincidence, causes the output in mode $1$ to become the same as
the input state in Eq.~(\ref{eq:scissin}) was. This is the case of successful
teleportation, which happens in the $1/9$ of the cases, regardless of the
input state in Eq.~(\ref{eq:scissin}).

In this section we have shown on an example, that the application of multiplet
concept in the description of photon number conservation can be indeed useful
in understanding operation of few-photon interference devices.

\section{OUTLINE OF AN $SU(3)$ THEORY OF TRITTERS}
\label{sect:tritters}

The question naturally arises, whether one can treat a passive linear optical
six-port or tritter in a similar manner to beam-splitters. Such devices can be
realized  either as a set of three coupled waveguides or as a
combination of beam-splitters and phase-shifters\cite{prl73_58}.
There are now three input and three output modes, thus both the input and
output can be regarded as a three-dimensional oscillator. Three-dimensional
oscillators are well known to possess $SU(3)$ symmetry. On the other hand, the
tritter can be described, similarly to the beam-splitter (\ref{eq:bsdef}), by
a unitary operator which is now element of $SU(3)$:
\begin{equation}
  \label{eq:tritterdef}
  \hat b_i=\sum_{j=1}^{2} U_{ij} \hat a_j, \quad i=1,2,3 \quad U\in SU(3),
\end{equation}
where $\hat a_i$-s and $\hat b_i$-s are the annihilation operators for the
input and output modes respectively. Thus we can follow the similar way, as in
the case of beam splitters: first we describe the bosonic realization of
$su(3)$ algebra and the structure of multiplets, then we introduce some
details of the tritter-transformation.

The lowest dimensional faithful representation of $su(3)$ algebra consists of
eight $3\times 3$ matrices $\frac{1}{2}\hat \lambda_i,\ i=1\ldots 8$, where
$\lambda _i$-s are the Gell-Mann matrices, explicitly:
\begin{eqnarray}
  \label{eq:selfrepr}
\hat  \lambda_1=\left( \matrix{ 0 & 1 & 0 \cr
                                  1 & 0 & 0 \cr
                                  0 & 0 & 0 \cr}\right),
\qquad
\hat   \lambda_2=\left( \matrix{ 0 & -i & 0 \cr
                                  i & 0 & 0 \cr
                                  0 & 0 & 0 \cr}\right),
\nonumber \\
\hat   \lambda_3= \left( \matrix{ 1 & 0 & 0 \cr
                                  0 & -1 & 0 \cr
                                  0 & 0 & 0 \cr}\right),
\qquad
\hat   \lambda_4=\left( \matrix{ 0 & 0 & 1 \cr
                                  0 & 0 & 0 \cr
                                  1 & 0 & 0 \cr}\right),
\nonumber \\
\hat   \lambda_5= \left( \matrix{ 0 & 0 & -i \cr
                                  0 & 0 & 0 \cr
                                  i & 0 & 0 \cr}\right),
\qquad
\hat   \lambda_6= \left( \matrix{ 0 & 0 & 0 \cr
                                  0 & 0 & 1 \cr
                                  0 & 1 & 0 \cr}\right),
\nonumber \\
\hat   \lambda_7= \left( \matrix{ 0 & 0 & 0 \cr
                                  0 & 0 & -i \cr
                                  0 & i & 0 \cr}\right),
\qquad
\hat   \lambda_8=\frac{1}{\sqrt3} \left( \matrix{ 1 & 0 & 0 \cr
                                         0 & 1 & 0 \cr
                                         0 & 0 & -2 \cr}\right).
\end{eqnarray}
The bosonic realization is given in a similar form as in
Eq.~(\ref{eq:su2_gen}) in the $su(2)$ case, namely for the input
operators: 
\begin{equation}
  \label{eq:su3real}
  \hat F_i=\frac{1}{2}\matrix{(\hat a_1^\dag &\hat a_2^\dag & \hat a_3^\dag )\cr & & \cr & &}
\left( \matrix{ & & \cr &\hat \lambda_i&\cr & & }\right)
\left(\matrix{\hat a_1\cr \hat a_2\cr \hat a_3}\right),
\end{equation}  
whereas the realization $\hat G_i,\ i=1\ldots 8$ for the output field
is defined in the same way with the operators $\hat b_k,\ k=1\ldots
3$.

Let us describe the multiplet structure at the input port. In order to do so,
we introduce some operators usually applied in this context:
\begin{eqnarray}
  \label{eq:opers}
  T_{\pm}=F_1\pm iF_2,\ U_\pm =F_6\pm iF_7,\nonumber \\
  V_\pm=F_4\pm iF_5,\ T_3=F_3,Y=\frac{2}{\sqrt 3}F_8.
\end{eqnarray}
The eigenvalues of the two commuting operators $T_3$ and $Y$ are applied for
labelling of the multiplets (Such as the third component of angular momentum
in the $SU(2)$ case). The others appear to be ``ladder-operators'', which
allow ``movements'' in the multiplets.  Before going into details, let us give
the explicit form of all the operators in for the input field: the generators
are
\begin{eqnarray}
  \label{eq:boson_gen}
  \hat F_1=\frac{1}{2}\left( \hat a_1^\dag \hat a_2 + \hat a_2^\dag\hat a_1 \right) \nonumber \\
  \hat F_2=\frac{i}{2}\left( \hat a_2^\dag \hat a_1 - \hat a_1^\dag\hat a_2 \right) \nonumber \\
  \hat F_3=\frac{1}{2}\left( \hat a_1^\dag \hat a_1 - \hat a_2^\dag\hat a_2 \right) \nonumber \\
  \hat F_4=\frac{1}{2}\left( \hat a_1^\dag \hat a_3 + \hat a_3^\dag\hat a_1 \right) \nonumber \\
  \hat F_5=\frac{i}{2}\left( \hat a_3^\dag \hat a_1 - \hat a_1^\dag\hat a_3 \right) \nonumber \\
  \hat F_6=\frac{1}{2}\left( \hat a_2^\dag \hat a_3 + \hat a_3^\dag\hat a_2 \right) \nonumber \\
  \hat F_7=\frac{i}{2}\left( \hat a_3^\dag \hat a_2 - \hat a_2^\dag\hat a_3 \right) \nonumber \\
  \hat F_8=\frac{1}{2\sqrt{3}}\left( \hat a_1^\dag \hat a_1 + \hat a_2^\dag\hat a_2- 2\hat a_3^\dag\hat a_3\right),
\end{eqnarray}
and the other operators:

\begin{eqnarray}
  \label{eq:boson_opers}
\hat   T_+=\hat a_1^\dag \hat a_2 \nonumber \\
\hat   T_-=\hat a_2^\dag \hat a_1 \nonumber \\
\hat   U_+=\hat a_2^\dag \hat a_3 \nonumber \\
\hat   U_-=\hat a_3^\dag \hat a_2 \nonumber \\
\hat   V_+=\hat a_1^\dag \hat a_3 \nonumber \\
\hat   V_-=\hat a_3^\dag \hat a_1 \nonumber \\  
\hat   T_3=\frac{1}{2}\left( \hat a_1^\dag \hat a_1 - \hat a_2^\dag\hat a_2 \right) \nonumber \\
\hat   Y=\frac{1}{3}\left(\hat a_1^\dag \hat a_1 + \hat a_2^\dag\hat a_2-2 \hat a_3^\dag\hat a_3\right).
\end{eqnarray}
Let us now examine the structure of the multiplets. There are two Casimir
operators of $SU(3)$, but they have a rather difficult structure, thus it is
conventional to use some other indexing of the multiplets.

In order to index states corresponding to the same multiplet, the eigenvalues
of the $\hat T_3$ and $\hat Y$ operator appear to be suitable, which are
linear combinations of the number operators $\hat a_1^\dag\hat a_1\ \hat
a_2^\dag\hat a_2$ and $\hat a_3^\dag\hat a_3$, and they commute with
them. Thus the eigenstates of these number-operators, the $|nlm\rangle$
Fock-states, are the eigenstates of $\hat T_3$ and $\hat Y$, with the
eigenvalues.
\begin{equation}
  \label{eq:Fockeig}
  \hat T_3|nlm\rangle=\frac{1}{2}(n-l)|nlm\rangle \quad,
 \hat Y|nlm\rangle=\frac{1}{3}(n+l-2m)|nlm\rangle. 
\end{equation}
Thus the multiplets can be visualized in the $T_3$--$Y$ plane.

\begin{figure}
 \begin{center}
   \epsfig{file=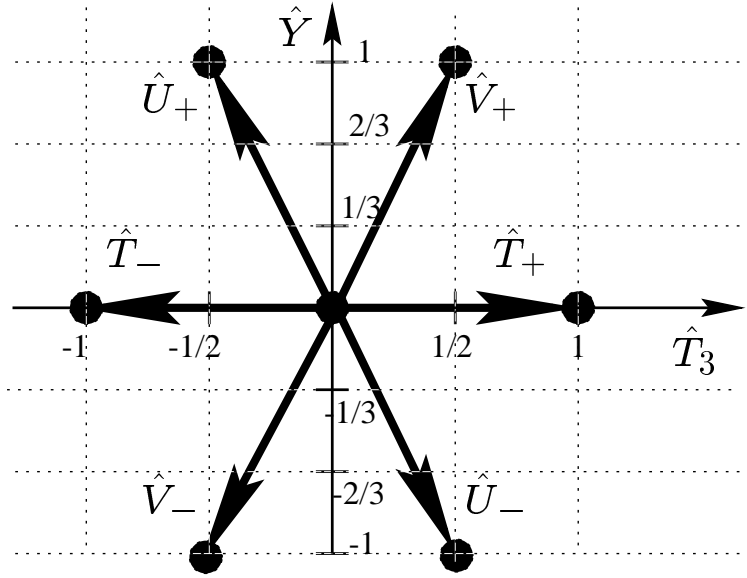,width=10cm}
\caption[oper]
{\label{fig:oper} 
Action of the ladder operators on the $T_3$ -- $Y$ plane 
}    
\end{center}
\end{figure}

If one of the elements of a given multiplet is known, the others can be
constructed using the $\hat U_\pm,\hat V_\pm,\hat T_\pm$ ladder-operators.
The action of these operators is shown in Fig. \ref{fig:oper}.
It can be seen, that the $su(3)$ algebra contains three $su(2)$
subalgebras symmetrically.

\begin{figure}
    \begin{center}
      \epsfig{file=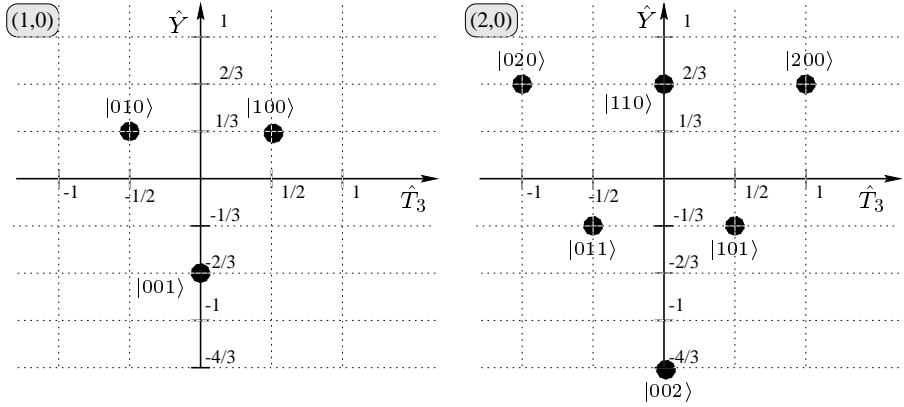,width=12cm}
\caption[multip]{
\label{fig:multip}
$SU(3)$-multiplets $(1,0)$ of one photon  and $(2,0)$ of two 
photons visualized on the $\hat T_3$ -- $\hat Y$ plane.
}
    \end{center}
\end{figure}

Generally, SU(3) multiplets are hexagon (truncated triangle) shaped (For a
simple explanation see e. g. in Ref.~\cite{Lipkin}).  Due to the difficult
structure of Casimir operators (not detailed here), usually the dimensions of
these polygons are used for indexing the states. The multiplet denoted by
$(\lambda,\mu)$ has
\begin{eqnarray}
  \label{eq:munu}
  \lambda=2 T_3, \ \mbox{at}\ Y=Y_{\mbox{\scriptsize{max}}} \\
  \mu=2 T_3, \ Y=Y_{\mbox{\scriptsize{min}}}.
\end{eqnarray}
\emph{Due to the additional symmetry of interchanging of bosons
  however, only the multiplets $(n,0)$ can be realized,
  where $n$ denotes the total number of the photons}.
Some of the multiplets are visualized on the $T_3$ -- $Y$
  plane in Fig. \ref{fig:multip}. 

Having described the multiplets, now let us turn our attention to the
description of the tritters alike that of beam splitters in
Ref.~\cite{pra40_1371}.  
The matrix $U$ describing the tritter in Eq.~(\ref{eq:tritterdef})
can be decomposed into a generalized ``Euler-angle''
parameterization defined in Ref.~\cite{physics9708015}:
\begin{equation}
  \label{eq:su3param}
U(\alpha,\beta,\gamma,\theta,a,b,c,\phi) = e^{(i\hat \lambda_3
  \alpha)} e^{(i\hat  \lambda_2 \beta)}e^{(i\hat  \lambda_3 \gamma)} e^{(i\hat
  \lambda_5 \theta)} e^{(i\hat  \lambda_3 a)} e^{(i\hat  \lambda_2 b)} 
e^{(i\hat  \lambda_3 c)} e^{(i\hat  \lambda_8 \phi)}.
\end{equation}
This expression enables us to connect a standard parameterization of
$SU(3)$ with the physical parameters of the actual realization of the
tritter, such as transmittivity and reflectivity coefficients of the
optical elements involved.

Similarly to the case of beam splitters, the action of a tritter may
be described as follows: the input of the tritter can be described by
the operators $\hat F_1\ldots \hat F_8$, which form a
$su(3)$-algebra. From these operators, all the others described in
Eqs. (\ref{eq:boson_opers}) can be expressed and used. The
beam-splitter transformation $U$ defined in Eq.
(\ref{eq:tritterdef}) turns these operators to operators
$\hat G_1\ldots \hat G_8$ describing the output modes. The
$\hat G$-s are formed from the $\hat b_1,\hat b_2,\hat b_3$
operators. The transformation $U$ is represented by the
\emph{adjoint representation} of $SU(3)$, which is a 8-parameter
subgroup of the group $SO(8)$ of 8-dimensional rotations:
\begin{equation}
  \label{eq:trafo}
\hat G_i = U\hat F_iU^\dag=\sum_{j=1}^8R_{ij}\hat F_j,
\quad R\in SO(8).
\end{equation}
The explicit form of the $R$ matrices is given in
Refs.~\cite{physics9803029,jphysa31_9255}.

\section{Conclusion}
\label{sect:concl}

In this paper we have revisited $SU(2)$-description of beam splitters. We have
presented an example of an optical setup in which only a few photon interfere,
and have shown how the multiplet picture can help in understanding of the
operation of such schemes.

We have outlined the $SU(3)$ description of optical tritters, describing
bosonic realization of $SU(3)$ and showing the structure of the
multiplets. The formalism suggested here maybe useful in applications of
tritters, e.g. in few-photon interference experiments. The situation in case
of tritters is similar to the case of the beam splitters, where 3 dimensional
rotations occur, but unfortunately 8 dimensional rotations cannot be
visualized.
 
\section*{Acknowledgements}

This work was supported by the Research Fund of Hungary
(OTKA) under contract No. T034484.


\end{document}